\documentstyle[preprint,aps]{revtex}
\input epsf
\begin{document}
\tighten
\title{
Probing $\bar u/\bar d$ Asymmetry in the Proton via\\
Quarkonium Production}

\author{J. C. Peng and D. M. Jansen}
\address{ Physics Division, Los Alamos National Laboratory\\
Los Alamos, New Mexico  87545}

\author{Y. C. Chen}
\address{Institute of Physics\\
Academia Sinica, Taipei, Taiwan}

\maketitle
\begin{abstract}
The sensitivity of proton-induced $J/\psi$ and $\Upsilon$ production to
the possible $\bar u/\bar d$
asymmetry in the nucleon is studied.  The ratio of the
quarkonium production cross sections at large $x_F$ on hydrogen
over deuterium targets, $\sigma (p+p) / \sigma (p+d)$, is
shown to be sensitive to this asymmetry.
Predictions of various theoretical models for this ratio are presented.
\end{abstract}

\vfill
\eject

\baselineskip 20pt

The New Muon Collaboration (NMC) reported~\cite{c1} the result of their
measurement of the Gottfried Sum~\cite{c2} in muon deep
inelastic scattering (DIS).
The Gottfried Sum is given as

\begin{eqnarray}
I_G(x_1, x_2;Q^2) = \int_{x_1}^{x_2} dx(F{_2}^{\mu p} (x,Q^2) - F{_2}^{\mu n}
(x,Q^2))/x,
\end{eqnarray}

\noindent where $F{_2}^{\mu p}$ and $F{_2}^{\mu n}$ are the
proton and neutron structure
functions measured in muon DIS.  Assuming charge symmetry for the parton
distributions in the proton and neutron, $F_2$ can be expressed in terms of
the valence and sea quark distributions of the proton.
The Gottfried Sum becomes

\begin{eqnarray}
I_G (x_1, x_2;Q^2) = {1 \over 3} \int_{x_1}^{x_2} dx [u_v (x, Q^2)
- d_v (x,Q^2)] + {2 \over 3} \int_{x_1}^{x_2} dx [\bar u
(x,Q^2) - \bar d (x,Q^2)].
\end{eqnarray}

\noindent The integral in the first term of (2) gives the difference
between the
number of the up and down valence quarks in the proton.  The second term in
(2) vanishes under the assumption of a $\bar u - \bar d$
flavor-symmetric sea for
the proton, and the Gottfried Sum Rule (GSR)~\cite{c2}

\begin{eqnarray}
I_G (0,1; Q^2) = {1 \over 3}
\end{eqnarray}

\noindent is obtained.
Based on their measurements of muon DIS on hydrogen and deuterium targets,
the NMC has
determined $I_G (0.004, 0.8; Q^2 = 4 GeV^2) = 0.227 \pm 0.007$~\cite{c1}.
Following an extrapolation to the unmeasured $x$-region, it was estimated
$I_G (0, 1; Q^2 = 4 GeV^2) = 0.240 \pm 0.016$, significantly different
from 1/3.

Many theoretical models [3-13] have been proposed to account for the apparent
violation of the GSR.  These models in general fall into two
categories.  Models in the first category~\cite{c3,c4} assume that the
valence quark distributions in the proton are sufficiently singular at
$x < 0.004$ such
that a large contribution to the Gottfried Sum occurs at
this region not probed by the NMC experiment.  Therefore,
the GSR is not violated and the assumption of
$\bar u/\bar d$ symmetry in
the proton remains valid in these models.  The other category of
theoretical models [5-13] interpret
the NMC result as evidence that the $\bar u$ and $\bar d$
distributions in the proton are different.  Some empirical expressions
for $\bar d(x) - \bar u(x)$ have been proposed~\cite{c6,c9,c13}.
Several recent sets of
parton distribution functions~\cite{c14,c15} explicitly
allow $\bar u/\bar d$ asymmetry
to account for the NMC result.
The origin of the enhancement of $\bar d$ over $\bar u$ in the proton has
been attributed to pion cloud~\cite{c5,c7,c8,c12}, diquark clustering
in the nucleon~\cite{c10}, as well as Pauli-blocking effect~\cite{c11}.

It has been proposed~\cite{c6,c16,c17}
that the Drell-Yan process provides an independent and
sensitive test of the possible  $\bar u/\bar d$ asymmetry in the
proton.  Using a proton beam and restricting the kinematic regime to
forward $x_F$ $(x_F > 0.2)$, it is straightforward to show that

\begin{eqnarray}
2 {\sigma_{DY}(p+p) \over \sigma_{DY}(p+d)} \simeq 1 -
{{\bar d_p (x) - \bar u_p (x)} \over {\bar d_p (x) + \bar u_p (x)}}.
\end{eqnarray}

\noindent A similar expression with a reduced sensitivity to the
$\bar u/\bar d$
asymmetry can also be obtained for the ratio of Drell-Yan cross sections
on neutron-rich targets versus isoscalar targets.  In fact, the E772
Drell-Yan data~\cite{c18} obtained with
tungsten and isoscalar targets have been compared with
predictions from various models.  More recently, the NA51 experiment~\cite{c19}
reported $2\sigma_{DY} (p+p)/\sigma_{DY} (p+d) = 0.91 \pm 0.02 \pm 0.02$
measured at 450 GeV near $x_F \simeq 0$ and $x = 0.18$. The NA51
result shows a significant asymmetry of $\bar u$ and $\bar d$ in the proton.
Another Drell-Yan experiment covering a
wider $x$ range $(0.05 < x < 0.3)$ at 800 GeV
has also been proposed~\cite{c20}.

In addition to the DIS and the Drell-Yan processes, there are other
interactions sensitive to the sea-quark distributions in the nucleon.  Of
particular interest is the hadronic production of $J/\psi$ and $\Upsilon$,
which can be detected in the same experiments designed to measure the
Drell-Yan process.  In this paper, the sensitivity of $J/\psi$ and
$\Upsilon$ production to the sea quark distributions in the nucleon is studied.

In contrast to the electromagnetic Drell-Yan process, hadronic $J/\psi$ and
$\Upsilon$ production involves strong interactions.  The
two processes responsible for producing
a pair of heavy quarks ($c \bar c$ and $b \bar b$)
are the $q \bar q$ annihilation and $gg$ fusion.
The cross sections for these QCD
subprocesses are~\cite{c21}

\begin{eqnarray}
\sigma (q \bar q \to Q \bar Q; m^2) & = & {{8 \pi \alpha_s^2}
\over{27m^6}}(m^2 +2m_Q^2) \lambda, \nonumber \\
\sigma (g g \to Q \bar Q; m^2) & = & {{\pi \alpha_s^2} \over{3m^6}}
 \{ (m^4 + 4m^2 m_Q^2 +m_Q^4) \ln({ {m^2+\lambda} \over {m^2-\lambda}})
- {1 \over 4}(7m^2 +31 m_Q^2)\lambda \},
\end{eqnarray}

\noindent where $m$ is
the invariant mass of the heavy quark pair ($Q \bar Q$), $m_Q$
is the heavy quark mass and $\lambda = (m^4 - 4m^2 m_{Q}{^2})^{1/2}$.
According to the QCD factorization theorem,
the differential cross section for producing a $Q \bar Q$ pair in hadronic
interaction is

\begin{eqnarray}
{{d \sigma} \over {dx_F d\tau}} = {{2\tau} \over
{(x_{F}{^2} + 4\tau^2)^{1/2}}} H_{PT} (x_1, x_2; m^2),
\end{eqnarray}

\noindent where
$x_1$, $x_2$ are the
fractional momenta carried by the projectile~($P$) partons and
target~($T$) partons, respectively,
$x_F = x_1 - x_2$ and $\tau^2 = m^2/S.$ $H_{PT}$ is the convolution of
parton cross sections and the parton distribution functions in the
projectile and target hadrons

\begin{eqnarray}
H_{PT} (x_1, x_2; m^2) = \,& & G_P(x_1) G_T(x_2) \, \sigma(gg \rightarrow
Q \overline{Q}; m^2) \, + \nonumber \\
 \sum_{\rm i=u,d,s} & & \bigl\{ q_P^{\,i}(x_1)
\overline{q}_T^{\,i}(x_2) + \overline{q}_P^{\,i}(x_1)
q_T^{\,i}(x_2) \bigr\} \,
\sigma(q \overline{q} \rightarrow Q \overline{Q}; m^2),
\end{eqnarray}

\noindent $G(x), q(x)$, and $\overline{q}(x)$ signify the gluon, quark, and
antiquark
distribution functions respectively.

To go from the production of
$Q \bar Q$ pair to the production of $Q \bar Q$ bound
states, we use the semi-local duality model~\cite{c22}.
In this model, the $Q \bar Q$ bound
state cross section is obtained by integrating the free
$Q \bar Q$ cross section
over $\tau$ from the $Q \bar Q$
threshold, $\tau_1 = 2m_Q/\sqrt s$, to the open
charm (or beauty) threshold, $\tau_2 = 2m_{D(B)}/\sqrt s$.  Hence

\begin{eqnarray}
d \sigma/dx_F (J/\psi,\Upsilon) = F \int_{\tau_1}^{\tau_2} 2\tau d\tau
H_{PT} (x_1, x_2; m^2)/(x_{F}{^2} + 4\tau^2)^{1/2},
\end{eqnarray}

\noindent where $F$ is the fraction of the $Q \bar Q$
bound state cross section
leading to $J/\psi (\Upsilon)$ production.

Despite its simplicity, the semi-local duality model is capable of
describing many features of hadronic $J/\psi$ and $\Upsilon$
productions~\cite{c23,c24,c25,c26}. In particular, the shape of
$d\sigma/dx_F$ as well as the dependences of the cross sections
on the beam energy and the incident-particle type, which are
very sensitive to the relative contributions
of $gg$ fusion and $q \bar q$ annihilation, are well reproduced.
The success of the
semi-local duality model suggests that it gives a reliable description
for the relative importance of these two processes.

\begin{figure}
\epsffile{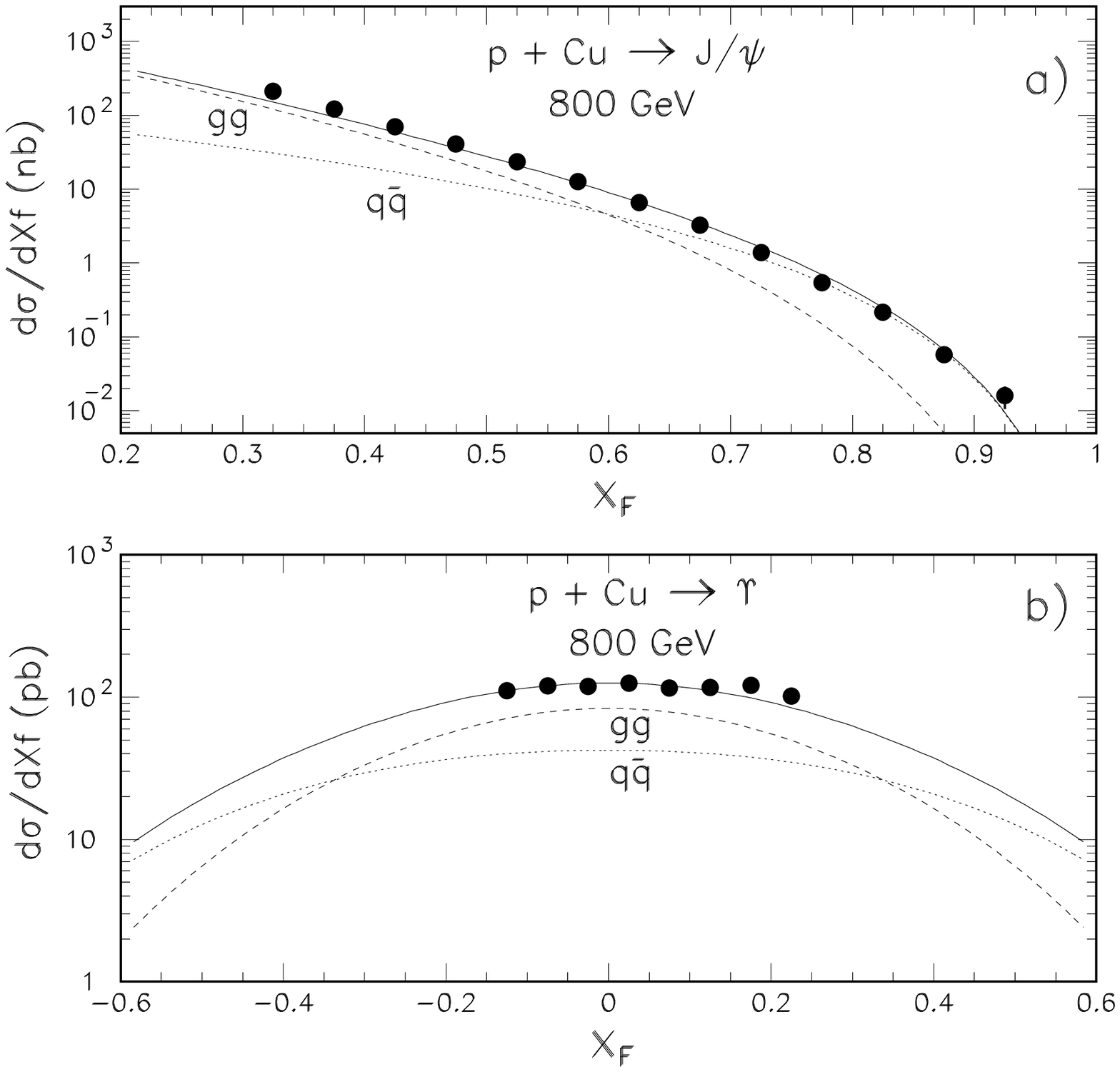}
\caption{
$J/\psi$ and $\Upsilon$ production cross sections in $ p + Cu$
interactions from References [26], [27]. The solid curves are calculations
using the semi-local duality model (Eq.(8)) and DO1.1 structure
functions [28].
The contributions from the $gg$ and $q \overline{q}$ annihilation
processes are shown as dashed and dotted curves respectively.}
\label{fig1}
\end{figure}

Figure~\ref{fig1} shows
the $d \sigma/dx_F$ data~\cite{c26,c27} for $J/\psi$ and $\Upsilon$
production using an 800 GeV proton beam.  Results of the semi-local
duality model calculations using Eq.(8) are shown as the solid curves.
The quark masses $m_c$ and $m_b$
were set at 1.5 GeV and 4.5 GeV, respectively, in the calculation.  The
recent structure function set, DO1.1~\cite{c28},
was used.  The shapes of the differential
cross sections are well described by the calculations, and $F$ is
determined to be 0.17 and 0.034, respectively, for $J/\psi$ and
$\Upsilon$ production.

The contributions from $gg$ fusion and
$q \bar q$ annihilation to the $J/\psi$
and $\Upsilon$ production cross sections are shown as the dashed and dotted
curves, respectively, in Figure~\ref{fig1}.
For $J/\psi$ production at this beam
energy, $gg$ fusion is the dominant process.  However, at
$x_F > 0.6$, $q \bar q$ annihilation process starts to
dominate. This reflects the fact that the gluon distribution drops
more rapidly at large $x$ than does
the quark distribution.  For $\Upsilon$
production, Figure~\ref{fig1}
shows that $q \bar q$ annihilation contributes $\sim$35\% of
the cross sections near $x_F=0$, and it is the dominant contribution
at large $x_F$.  Since $\Upsilon$ is roughly three
times more massive than $J/\psi$,
production of $\Upsilon$ is more sensitive to structure functions at
large $x$, where gluon ceases to dominate the parton densities.
This accounts for the relative importance of $q \bar q$ annihilation for
$\Upsilon$ production.

For $J/\psi$ and $\Upsilon$ production in $p-p$ interactions, the
cross section from $u \bar u$ and $d \bar d$
annihilation, $\sigma(p + p)$, is
proportional to

\begin{eqnarray}
u_p (x_1) \bar u_p (x_2) + \bar u_p (x_1) u_p (x_2)
+ d_p (x_1) \bar d_p (x_2) + \bar d_p (x_1) d_p (x_2).
\end{eqnarray}

\noindent For $p-d$ interactions, the corresponding cross section,
$\sigma(p + d)$, is

\begin{eqnarray}
{ u_p (x_1) (\bar u_p (x_2) + \bar u_n (x_2)) + \bar
u_p (x_1) (u_p (x_2) + u_n (x_2))} \atop {+ d_p (x_1) (\bar d_p (x_2) + \bar
d_n (x_2)) + \bar d_p (x_1) (d_p (x_2) + d_n (x_2)).}
\end{eqnarray}

\noindent In Eq.(10) the deuteron
parton distributions are taken as the sum of the proton~($p$) and
neutron~($n$) distributions.
Charge symmetry requires $u_p (x) = d_n(x), \bar u_p(x) = \bar d_n(x)$, etc.
Expression (10) now becomes

\begin{eqnarray}
{ u_p (x_1) (\bar u_p (x_2) + \bar d_p (x_2)) + \bar u_p
(x_1) (u_p (x_2) + d_p (x_2))} \atop {+ d_p (x_1) (\bar d_p (x_2) + \bar u_p
(x_2)) + \bar d_p (x_1) (d_p (x_2) + u_p (x_2))}
\end{eqnarray}

\noindent For a flavor-symmetric sea, $\bar u_p$ = $\bar d_p$, Eqs.
(9) and (11) give $\sigma (p+d) = 2 \sigma (p+p)$. This relation is also
valid for contributions from the $s \bar s$ annihilation and $gg$ fusion.
Therefore, the ratio $R(x_F)$, defined as

\begin{eqnarray}
R(x_F) = 2 \ {{d\sigma/dx_F \ (p + p \to J/\psi (\Upsilon))} \over
{d\sigma/dx_F (p + d \to J/\psi (\Upsilon))}}
\end{eqnarray}

\noindent would be equal to 1 for all models which assume $\bar u_p = \bar
d_p$.
On the other hand, $R(x_F)$ could deviate significantly
from 1 if $\bar u_p \ne \bar d_p$, especially at large $x_F$ where
$q \bar q$ annihilation has a dominant contribution.

\begin{figure}
\epsffile{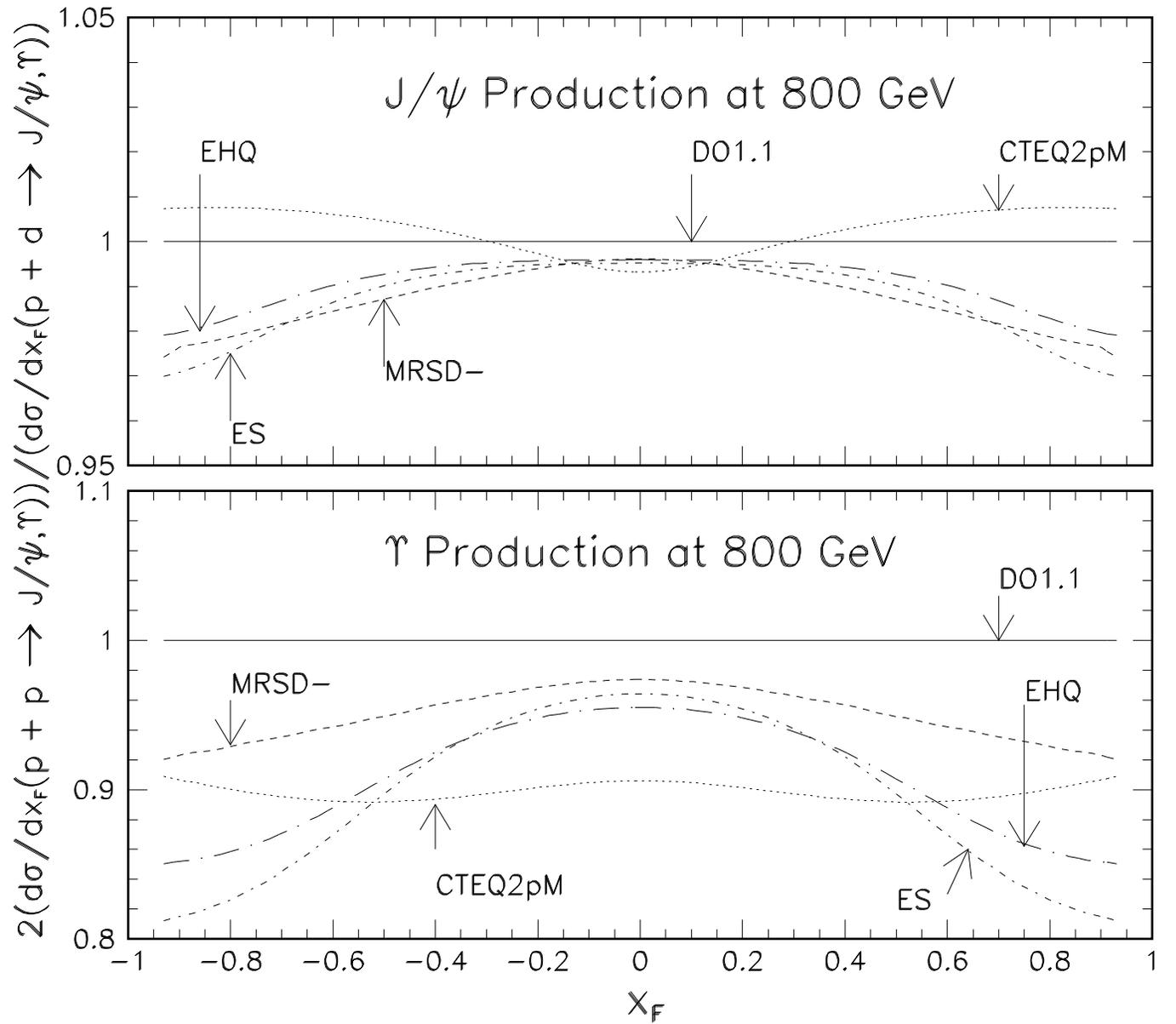}
\caption{
Predictions of $R(x_F)$ for $J/\psi$ and $\Upsilon$ production
at 800 GeV using various structure function sets in the semi-local
duality model.}
\label{fig2}
\end{figure}

Figure~\ref{fig2} shows the predictions of $R(x_F)$ for $J/\psi$ and $\Upsilon$
production with 800 GeV proton beam.  Five different structure function
sets have been used in the calculations.  The solid curves correspond to
the $\bar u/\bar d$
symmetric DO1.1 structure functions.  As discussed earlier,
$R(x_F)$ is identically one in this case.  The dashed and dotted curves are
results for the $\bar u/\bar d$
asymmetric structure function sets MRSD-$^\prime$ and
CTEQ2pM, respectively.  These two structure function sets were obtained from
recent global fits to Drell-Yan and DIS data including
the NMC result~\cite{c14,c15,c29}.
Finally, the calculations using the parameterization of $\bar d(x) - \bar u(x)$
given by Ellis and Stirling~\cite{c6}
and Eichten {\it et al.}~\cite{c13} are also shown in Figure~\ref{fig2}.
The parameterizations for $\bar d(x)-\bar u(x)$ were given at a fixed value
of $Q^2$ at 4 GeV$^{2}$. The $Q^2$-dependence was taken into account in the
calculation by using the Altarelli-Parisi evolution
equation for flavor non-singlet structure functions~\cite{c30}.

\begin{figure}
\epsffile{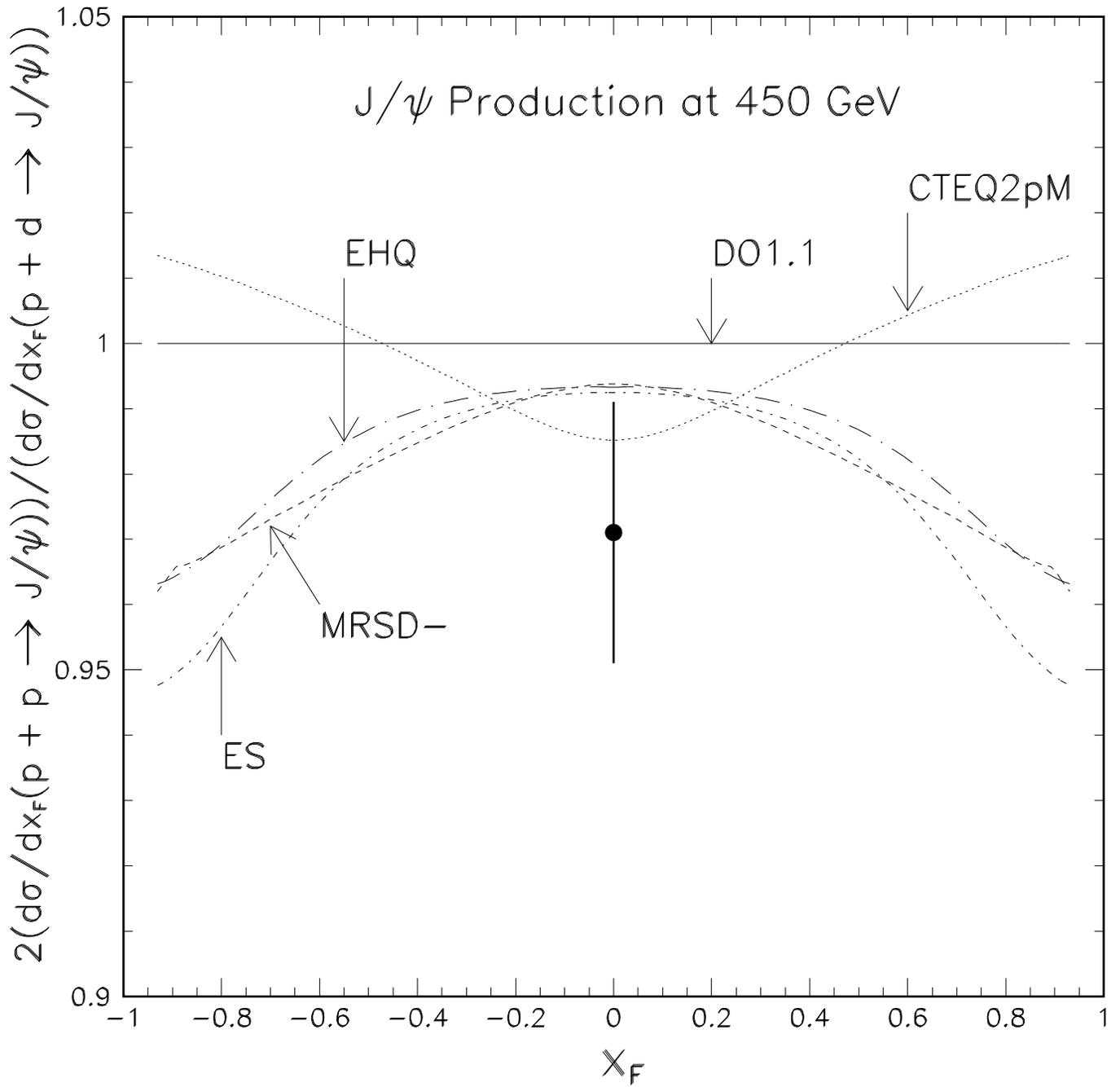}
\caption{
Predictions of $R(x_F)$ for $J/\psi$ production at 450 GeV
using various structure function sets in the semi-local duality model.
The data from NA51 [19] is also shown.}
\label{fig3}
\end{figure}

Figure~\ref{fig2} shows
that the measurement of $R(x_F)$ for $\Upsilon$ production
at 800 GeV provides a sensitive test of models which have different
descriptions on the $\bar u/\bar d$
symmetry in the proton.  In comparison,
$J/\psi$ production is much less sensitive to the sea quark distributions
due to the dominance of $gg$ fusion at this energy.  However, at lower beam
energies, $J/\psi$ production is more sensitive to the sea quark
distributions as it probes a larger $x$ region where $q \bar q$
annihilation starts to be important.  This is illustrated in
Figure~\ref{fig3} which
shows the predictions of $R(x_F)$ for $J/\psi$ production using 450 GeV
proton beam.  The recent result from NA51~\cite{c19}
is also compared with the
predictions.  While the predictions are consistent with the data, it is
important to extend the measurements to larger $x_F$ values.  A measurement
of $R(x_F)$ for $J/\psi$ production at even lower beam energies
would also be of interest.

It should be mentioned that a more elaborate scheme for calculating
hadronic $J/\Psi$ and $\Upsilon$ production, the so-called color-singlet
model, has also been proposed~\cite{c31,c32}. This model includes relevant
$gg$, $gq$, $g \bar q$, and $q \bar q$ processes up to order of $\alpha_s^3$,
and is capable of predicting the absolute production cross sections. It
would be very interesting to compare the predictions of the semi-local
duality model and the color-singlet model with the experimental data.

In conclusion, proton-induced $J/\psi$ and $\Upsilon$ production offers
an independent means to examine the possible $\bar u/\bar d$
asymmetry in the proton.  Measurement of the cross
section ratios on hydrogen and
deuterium targets in the forward $x_F$ region would provide a sensitive test
of various theoretical models on the sea quark distributions of the proton.

\vfill
\eject

\end{document}